\documentclass[conference,9pt]{IEEEtran}

\usepackage{cite}
\usepackage{amsmath}
\usepackage{graphicx}
\usepackage{booktabs}
\usepackage{makecell}
\usepackage{balance}
\usepackage{flafter}
\usepackage{algorithm}
\usepackage{algpseudocode}
\usepackage[hidelinks]{hyperref}

\newcommand{\ours}{\textsc{NEAT}}
\newcommand{\wap}{Transport-Informed Placement}
\newcommand{\ppr}{Partial-Progress Routing}

\begin{document}
\bstctlcite{IEEEexample:BSTcontrol}

\title{Efficient AOD-Transport Compilation for Neutral-Atom Arrays}

\author{\IEEEauthorblockN{
Haydn Cheng\textsuperscript{1}, Zhiding Liang\textsuperscript{2,3,\textdagger}}
\IEEEauthorblockA{
\textsuperscript{1}University of Oxford\quad
\textsuperscript{2}The Chinese University of Hong Kong\\
\textsuperscript{3}State Key Laboratory of Quantum Information Technologies and Materials (CUHK)\\
pak.cheng@magd.ox.ac.uk\quad
\textdagger{}Corresponding author: zliang@cse.cuhk.edu.hk}
\vspace{-0.3in}}

\maketitle

\begin{abstract}
Neutral-atom quantum computers execute two-qubit gates by shuttling atoms between storage and entanglement zones with acousto-optic deflectors (AODs). The efficiency of these movements determines circuit fidelity via transport overhead, idle-atom crosstalk, and atom-transfer errors. Additionally, AOD hardware constraints limit parallel atom movements, so efficient qubit placement and routing are essential.

We present \ours{} (\textbf{N}eutral-atom \textbf{E}fficient \textbf{A}OD-\textbf{T}ransport compiler). Its transport-informed placement evaluates each decision by its incremental equivalent time, a single cost combining transport time, atom-transfer loss, and crosstalk. Its partial-progress routing allows conflicting moves to advance concurrently via intermediate waypoints. \ours{} also schedules gates to maximize parallelism and uses a staggered atom array whose trap geometry preemptively avoids AOD violations without extra moves.

On standard benchmark circuits, \ours{} outperforms state-of-the-art (SOTA) compilers in fidelity, circuit duration, and compilation time. On scalability benchmarks of up to 1{,}000 qubits, it shortens circuits by \textbf{1.9$\times$}, reduces the negative log-fidelity by \textbf{1.5$\times$}, and compiles \textbf{6.6$\times$} faster than SOTA compilers on average. Overall, \ours{} is a lightweight yet effective compiler for efficient and scalable neutral-atom quantum computing. The code is available at \url{https://github.com/HaydnCheng77/NEAT}.

\end{abstract}

\section{Introduction}

As quantum computing progresses from the Noisy Intermediate-Scale Quantum (NISQ) era toward the early fault-tolerant regime~\cite{preskill2018}, many qubit modalities are being pursued, \textit{e.g.,} superconducting circuits, trapped ions, photonic systems, and quantum dots. Among these, neutral-atom quantum computers are promising due to their scalability, with demonstrated arrays of over 6{,}100 atomic qubits~\cite{manetsch2024}.

In zoned neutral-atom architectures~\cite{bluvstein2023}, idle atoms are kept in a \emph{storage zone}, and atoms due for a two-qubit gate are transported by acousto-optic deflectors (AODs)~\cite{bluvstein2022} to an \emph{entanglement zone}, where a global Rydberg beam applies CZ gates. This reconfigurability allows any two atoms to be entangled directly, giving all-to-all connectivity at a scale no other platform matches~\cite{kim2023utility,guo2024ions,manetsch2024}. However, such connectivity comes at a cost. Every atom transfer between a static trap and an AOD trap costs time and fidelity, every idle atom left in the entanglement zone suffers crosstalk, and moving atoms back and forth adds decoherence. More importantly, AOD transport must obey hardware rules that highly constrain parallel movements (Sec.~\ref{sec:aod-model}).

A neutral-atom compiler must therefore solve two tightly coupled problems. \emph{Qubit placement} decides where each atom sits at each stage so that the required moves can be routed efficiently. \emph{Qubit routing} then decides how to realize these moves efficiently under the AOD rules. Earlier compilers~\cite{zac,powermove} treat placement and routing as separate steps, so they can produce conflicting placements the router cannot execute efficiently. More recent compilers~\cite{stade2025routing,stade2025search,zap} account for AOD conflicts during placement, but they approximate these conflicts coarsely, estimating them with distance-based proxies and hand-tuned weights. Moreover, existing routers defer a move whenever it conflicts with the moves already selected for the same round, which lengthens the circuit and therefore adds decoherence to every qubit.

In this paper, we present \ours{}, a zoned neutral-atom compiler addressing these limitations through three main contributions:
\begin{itemize}
  \item We propose \textbf{\wap{}}, which combines the infidelity channels into a unified equivalent-time cost heuristic and evaluates each decision by its incremental cost.
  \item We introduce \textbf{\ppr{}}, which exploits AOD parallelism by allowing conflicting atom movements to make partial progress through temporary intermediate placements.
  \item Beyond placement and routing, we design a staggered array to reduce AOD violations and a parallelism-enhancing schedule.
\end{itemize}

\section{Background}
\label{sec:background}

\subsection{Neutral-Atom Quantum Computers}
\label{sec:zoned}

Neutral-atom quantum computers encode qubits in internal states of atoms (\textit{e.g.,} Rb~\cite{bluvstein2022} and Cs~\cite{manetsch2024}). Single-qubit gates are driven by Raman lasers~\cite{levine2022raman} and can run in parallel~\cite{bluvstein2023}. Two-qubit CZ gates use the Rydberg blockade~\cite{jaksch2000,saffman2010,saffman2016}. A Rydberg laser pulse entangles only atoms closer than the blockade radius $R_B$. In~\cite{bluvstein2023}, $R_B = 4.3\,\mu$m.

Static atoms are held in optical tweezers generated by a spatial light modulator (SLM), which can create arbitrary trap geometries~\cite{nogrette2014} with minimum spacing set by optical resolution~\cite{sortais2007optics,tan2024olsq}. Zoned architectures~\cite{bluvstein2023} divide these traps into two zones. The \emph{entanglement zone} arranges traps into Rydberg pairs, with the two sites of a pair closer than $R_B$ and distinct pairs far apart, so a global Rydberg pulse applies CZ gates to all pairs simultaneously. The \emph{storage zone} is a dense array of traps, typically $2$--$6\,\mu$m apart~\cite{bluvstein2023}, that holds idle qubits more than $R_B$ from the entanglement zone to avoid crosstalk.

\subsection{AOD Movement Model}
\label{sec:aod-model}
\label{sec:transport}

\begin{figure}[!tbp]
\centering
\includegraphics[width=\columnwidth]{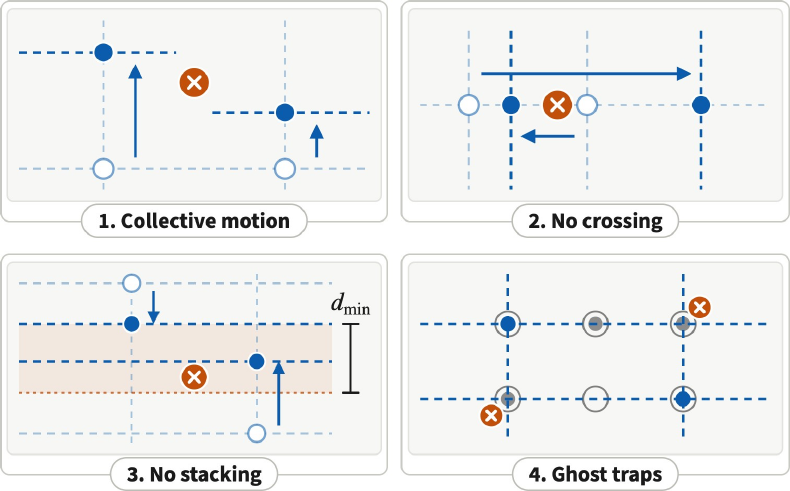}
\caption{Four AOD movement rules. Blue (gray) circles represent the AOD (SLM)-trapped atoms. Red crosses mark the violations of each rule.}
\label{fig:aod-model}
\end{figure}

Atoms move between SLM traps in mobile traps generated by two orthogonal AODs. Each AOD tone sets one column ($x$) or row ($y$), and a trap forms at every intersection of an active row and column~\cite{tan2024olsq}. To transport an atom, the AOD activates its row and column, sweeps them to the destination, and deactivates either line to release it~\cite{tan2024olsq}. As in prior compilers~\cite{zac,powermove,zap}, a move over distance $d$ takes \(t(d) = \sqrt{d/a}\), where $a = 2750\,\mathrm{m\,s^{-2}}$ is the maximum acceleration that maintains coherent transport~\cite{bluvstein2022}. A set of AOD-compatible moves performed in parallel, a \emph{wave}, takes as long as its longest move and must satisfy the four rules in Fig.~\ref{fig:aod-model}.

\begin{figure*}[!tbp]
\centering
\includegraphics[width=\textwidth]{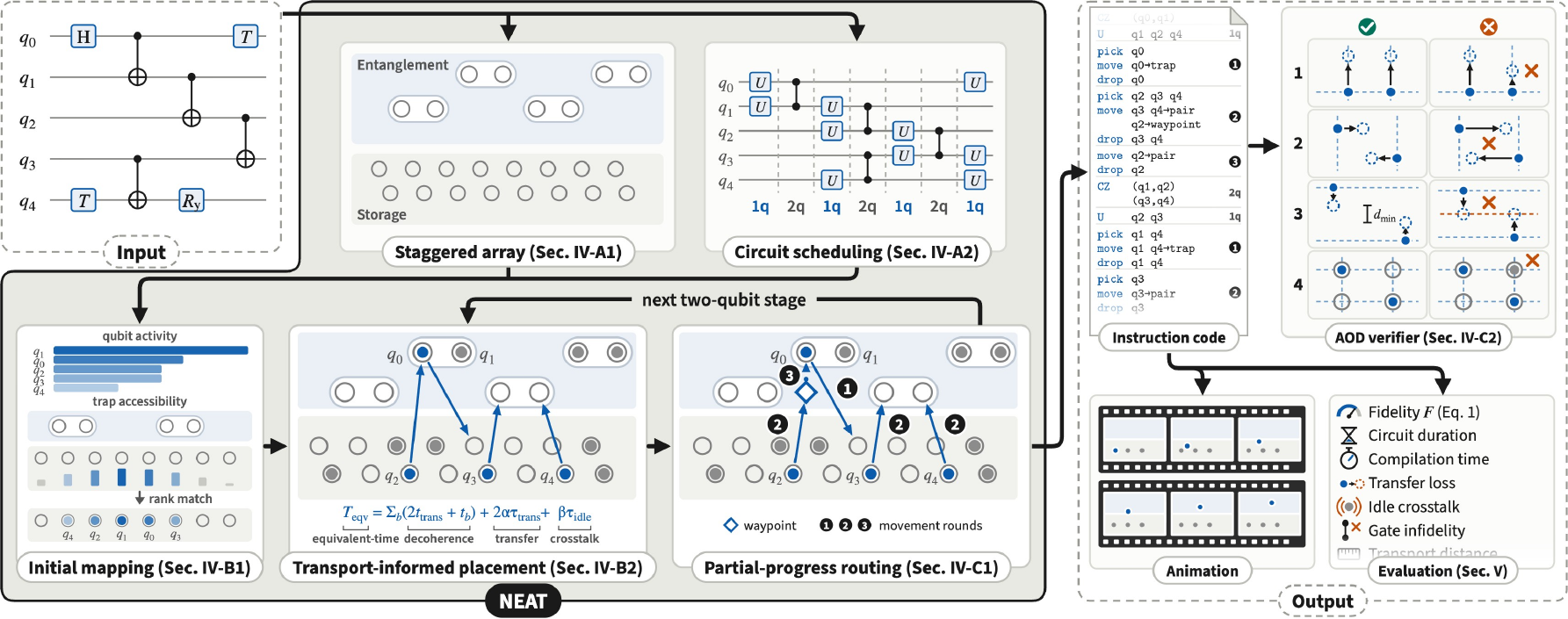}
\caption{A high-level overview of the end-to-end compilation flow of \ours{}.}
\label{fig:pipeline}
\end{figure*}

\textbf{Rule~1 (Collective motion).} Atoms sharing an AOD row (column) must undergo the same vertical (horizontal) displacement.

\textbf{Rule~2 (No crossing).} Active AOD rows (columns) must keep their top-to-bottom (left-to-right) order to avoid heating and atom loss~\cite{bluvstein2022}.

\textbf{Rule~3 (No stacking).}
Distinct active rows (columns) must stay at least $d_{\min}=2\,\mu\mathrm{m}$ apart~\cite{tan2024olsq}. Thus, two atoms on distinct rows (columns) can never be moved onto the same final row (column).

\textbf{Rule~4 (Ghost traps).}
Activating the AOD rows and columns of atoms at $(x_1,y_1)$ and $(x_2,y_2)$ also activates the corner traps at $(x_1,y_2)$ and $(x_2,y_1)$, and any atoms there are picked up as well.

\subsection{Fidelity Model}
\label{sec:fidelity-model}

We model circuit fidelity as the product of five independent factors,
\begin{equation}
F = F_{\mathrm{2q}} \cdot F_{\mathrm{1q}} \cdot F_{\mathrm{trans}}
    \cdot F_{\mathrm{idle}} \cdot F_{\mathrm{coh}}.
\label{eq:fidelity}
\end{equation}

The gate factors are $F_{\mathrm{2q}} = f_{\mathrm{2q}}^{\,N_{\mathrm{2q}}}$ and $F_{\mathrm{1q}} = f_{\mathrm{1q}}^{\,N_{\mathrm{1q}}}$, where $N_{\mathrm{2q}}$ and $N_{\mathrm{1q}}$ are the numbers of two- and single-qubit gates, with fidelities $f_{\mathrm{2q}} = 0.995$ and $f_{\mathrm{1q}} = 0.9997$~\cite{evered2023}. They are the same for all compilers. Atom transfers give $F_{\mathrm{trans}} = f_{\mathrm{trans}}^{\,N_{\mathrm{trans}}}$, where $N_{\mathrm{trans}}$ is the number of pick-ups and drop-offs and $f_{\mathrm{trans}} = 0.999$~\cite{bluvstein2023}. Global Rydberg pulses cause crosstalk on idle atoms in the entanglement zone, giving $F_{\mathrm{idle}} = f_{\mathrm{idle}}^{\,N_{\mathrm{idle}}}$, where $N_{\mathrm{idle}}$ counts idle-atom exposures and $f_{\mathrm{idle}} = 0.9975$~\cite{zac}. We approximate the decoherence factor as
\begin{equation}
F_{\mathrm{coh}} =
\exp\left(-\frac{n T_{\mathrm{total}}}{T_2}\right),
\label{eq:coherence}
\end{equation}
where $n$ is the number of qubits, $T_2 = 1.5\,$s~\cite{bluvstein2022} is the coherence time, and $T_{\mathrm{total}}$ is the circuit duration, the sum of the times of all operations: $t_{\mathrm{1q}} = 52\,\mu$s~\cite{zac} for one 1q stage, $t_{\mathrm{2q}} = 0.27\,\mu$s~\cite{bluvstein2023} for one 2q stage, $t_{\mathrm{trans}} = 15\,\mu$s~\cite{zac} for one atom transfer round, and $t(d)$ (Sec.~\ref{sec:transport}) for one movement round, set by its longest move.

\section{Motivation}
\label{sec:motivation}

\subsection{Related Work}
\label{sec:related}

Early compilers target single-zone architectures~\cite{enola,geyser,q-pilot,atomique,dasatom,parallax}, while recent ones focus on zoned architectures~\cite{nalac,zac,powermove,zap,stade2025routing,stade2025search,kong2026directional}.

PowerMove~\cite{powermove} and ZAC~\cite{zac} avoid crosstalk on idle qubits by returning qubits to the storage zone unless they participate in an immediate gate. Their placement strategies, however, prioritize movement distance and qubit reuse, without considering AOD compatibility. As a result, their atom transports are often heavily serialized~\cite{stade2025routing}.

To address this, MQT QMAP~\cite{stade2025routing} incorporates movement compatibility into its objective and uses A* search to find low-cost placements. Its follow-up work~\cite{stade2025search} replaces the memory-intensive A* search with Iterative Diving Search to improve scalability. ZAP~\cite{zap} instead adds a weighted AOD-conflict penalty to its heuristic in a deterministic, single-pass flow that avoids iterative search.

For routing, these compilers largely share a common framework: atom movements are grouped and executed simultaneously by AOD compatibility. PowerMove groups movements greedily after sorting them by distance~\cite{powermove}. MQT~\cite{stade2025routing} uses a similar algorithm, and introduces relaxed routing~\cite{stade2025search}, which partly relaxes the no-crossing rule (Rule~2). ZAC adopts Enola's grouping based on maximal independent sets (MIS)~\cite{enola,zac}, whereas ZAP combines its safe-movement selection with greedy MIS-based move grouping.

\subsection{Proposed Solution}
\label{sec:proposed}

Building on prior work, we propose \ours{}. Its transport-informed placement evaluates each decision by its incremental equivalent time, a cost heuristic that unifies transport decoherence, atom-transfer errors, and idle-atom crosstalk over the stage's planned AOD waves.

We also reconsider how conflicting moves are executed. Apart from short offset moves that avoid ghost traps (Rule~4, Sec.~\ref{sec:aod-model})~\cite{zac,zap,stade2025search} and MQT's relaxed routing in specific cases~\cite{stade2025search}, existing compilers mostly serialize them. Partial-progress routing instead moves a conflicting atom as far toward its destination as the AOD rules allow and holds it there. Once the atoms blocking it are released, it completes the remaining distance. Moves that would otherwise be deferred therefore make progress without violating AOD constraints.

\ours{} also schedules gates to enhance parallelism and replaces the conventional rectangular atom array with a staggered one that geometrically reduces ghost-trap violations. A built-in AOD verifier ensures that every compiled move is physically executable.

\section{\ours{} Compiler}
\label{sec:neat}

Fig.~\ref{fig:pipeline} shows an overview of \ours{}: the circuit is transpiled and scheduled into stages on a staggered array, and each stage is placed and routed. The resulting instruction code is evaluated and animated, and a verifier checks it against the four AOD rules of Fig.~\ref{fig:aod-model}.

\subsection{Preprocessing}
\label{sec:pipeline}

\subsubsection{Staggered array}
\label{sec:array}
Before placement begins, \ours{} builds a staggered array sized to the circuit, with $n$ Rydberg pairs and $1.5n$ storage traps. Most prior compilers use rectangular SLM trap arrays, whose shared rows and columns make them prone to ghost pick-ups (Rule~4, Sec.~\ref{sec:aod-model}). To avoid lifting unintended atoms, compilers are forced to  pick up atoms sequentially, with small offset moves between pick-ups~\cite{zap,zac,stade2025search}. However, these moves and atom-transfer steps lengthen the circuit and increase decoherence.

\begin{figure}[!tbp]
\centering
\includegraphics[width=\columnwidth]{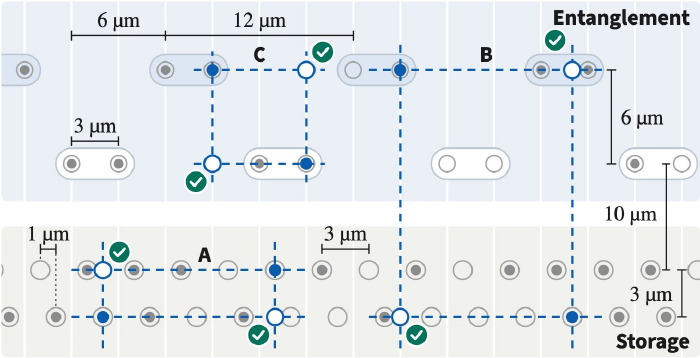}
\caption{Schematic of a staggered atom array. Storage rows alternate $x \equiv 0$ and $x \equiv 1 \pmod 3$. Entanglement sites sit at $x \equiv 2 \pmod 3$ in two classes offset by $6\,\mu$m. (A): two storage atoms in different classes. (B): storage and entanglement atoms. (C): two entanglement atoms in different classes.}
\label{fig:stagger}
\end{figure}

We instead use a staggered array to prevent ghost pick-ups through its geometry (Fig.~\ref{fig:stagger}). The storage zone alternates between two row classes: one has traps at $x \equiv 0 \pmod 3$ (in $\mu$m), and the other is shifted laterally by $1\,\mu$m to $x \equiv 1 \pmod 3$. Rows of different classes never share a column, so picking up two atoms from different classes lifts no unintended atom (Fig.~\ref{fig:stagger}A). However, columns closer than $d_{\min}$ must still be picked up in separate rounds (Rule~3).

The entanglement zone follows similar principles. Its traps satisfy $x \equiv 2 \pmod 3$, so no column passes through a storage trap, and an entanglement atom and a storage atom can be picked up together without a ghost pick-up (Fig.~\ref{fig:stagger}B) if their columns are at least $d_{\min}$ apart. Within the zone, two row classes offset laterally by $6\,\mu$m prevent ghost pick-ups between atoms in different classes (Fig.~\ref{fig:stagger}C).

The layout is realizable through programmable SLM phase patterns (Sec.~\ref{sec:zoned}). The finite waist of the trapping beams, around $1\,\mu$m~\cite{sortais2007optics}, limits how small the lateral shift between classes can be. 
This arrangement also raises the minimum inter-pair trap distance from $6\,\mu$m in a rectangular array to $\sqrt{6^2 + 3^2} \approx 6.7\,\mu$m without reducing trap density, further suppressing inter-pair interactions.

\subsubsection{Circuit scheduling}
\label{sec:schedule}

\begin{figure}[!tbp]
\centering
\includegraphics[width=\columnwidth]{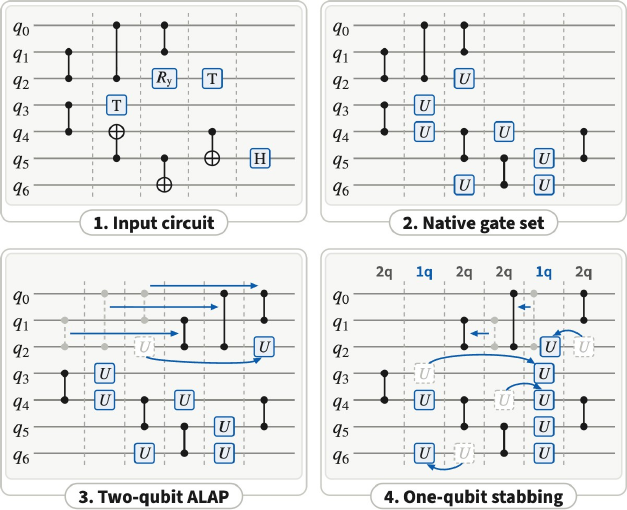}
\caption{Circuit scheduling. After transpilation to native gates, CZ gates are scheduled ALAP, and single-qubit gates are grouped into the fewest 1q-stages.}
\label{fig:schedule}
\end{figure}

Fig.~\ref{fig:schedule} illustrates circuit scheduling. We first transpile the input circuit into CZ gates and arbitrary single-qubit rotations ($U$) using Qiskit at optimization level~3~\cite{qiskit_optimization}. We then schedule the CZ gates into two-qubit stages (2q-stages) as late as possible (ALAP) while preserving circuit dependencies. Like as-soon-as-possible (ASAP) scheduling, ALAP achieves the minimum number of 2q-stages, set by the longest dependency path. However, ALAP also delays gates with slack, keeping their atoms in storage longer and thus reducing their exposure to Rydberg crosstalk.

With the 2q-stages fixed, we schedule single-qubit rotations into single-qubit stages (1q-stages). Each rotation can go in any gap between its preceding and succeeding CZ gates on its qubit. Since rotations on different qubits can run simultaneously, our objective is to minimize the number of 1q-stages to reduce circuit duration. This is the interval-stabbing problem, where each rotation has a range of valid gaps (intervals) and the goal is to select as few 1q-stages (stabs) as possible. A greedy algorithm solves it optimally: we visit the rotations in increasing order of their last gap, and whenever no 1q-stage yet falls in a rotation's range, we create one at that gap. Each rotation then runs in the latest 1q-stage within its range.

\subsection{Qubit Placement}
\label{sec:tip}

\subsubsection{Initial placement}
\label{sec:init-mapping}

Our initial placement matches the most active qubits with the storage traps that have the best access to the entanglement zone. Qubit activity is a stage-weighted gate count: each gate of a qubit at stage $s$ contributes $1/(s+1)$, so qubits that gate often and early rank highest. Trap accessibility is the mean distance to every entanglement site. Sorting both lists and matching them rank for rank places the busiest qubits closest to their future gates.

\begin{figure*}[!tbp]
\centering
\includegraphics[width=\textwidth]{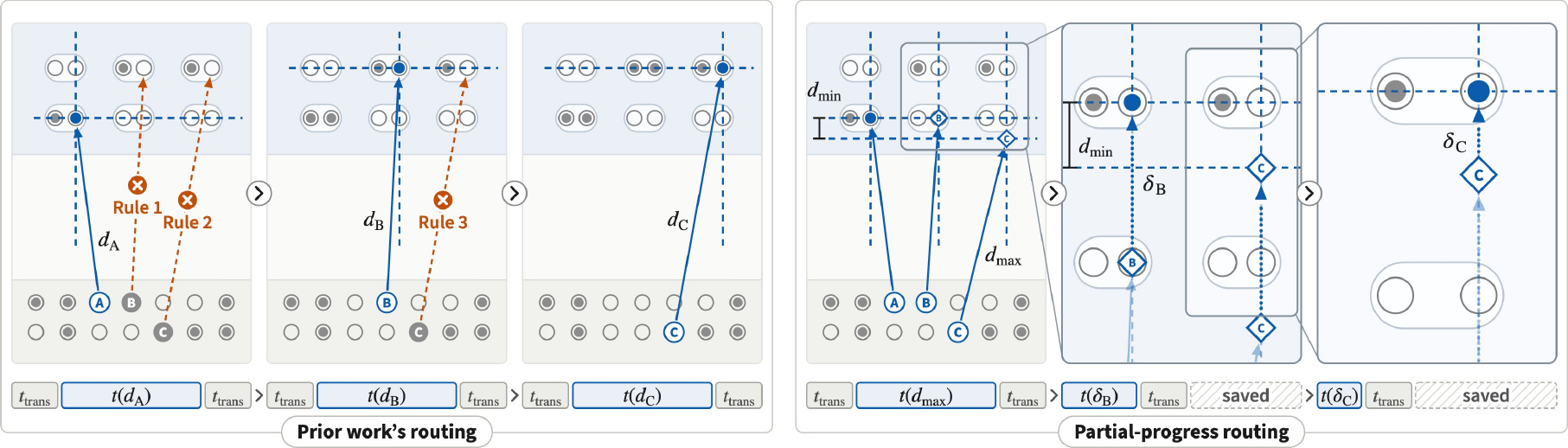}
\caption{Prior work's routing (left) moves A, B, and C in three waves due to AOD conflicts. Partial-progress routing (right) moves A, B, and C together by introducing waypoints (diamonds). Durations are compared at the bottom, and the dashed hatched blocks show the time saved.}
\label{fig:ppr}
\end{figure*}

\subsubsection{Stage placement}
\label{sec:stage-placement}

For each two-qubit stage $\ell$, the placer decides which idle qubits to evacuate from the entanglement zone (and to which storage traps), and which Rydberg pair each two-qubit gate uses. It evaluates both decisions by one unified cost, the equivalent time of the moves the router must later perform for the stage.

To estimate this cost, the placer predicts how the router will group the stage's moves into AOD-compatible waves (Sec.~\ref{sec:aod-model}). Each wave $b$ spends $2t_{\mathrm{trans}}$ on pick-up and drop-off and $t_b$ on transport, where $t_b$ is the duration of its slowest move. To also account for the fidelity lost due to atom transfers and idle-atom crosstalk, we write the total cost of a stage whose moves form the waves $B$ as an equivalent time,
\begin{equation}
\begin{aligned}
T_{\mathrm{eqv}}(B) &= \sum_{b \in B} \bigl(2t_{\mathrm{trans}} + t_b\bigr) + 2\alpha\,\tau_{\mathrm{trans}} + \beta\,\tau_{\mathrm{idle}},\\
\tau_{\mathrm{trans}} &= -\frac{T_2 \ln f_{\mathrm{trans}}}{n}, \qquad \tau_{\mathrm{idle}} = -\frac{T_2 \ln f_{\mathrm{idle}}}{n},
\end{aligned}
\label{eq:stage-cost}
\end{equation}
where $\alpha$ is the number of moving atoms and $\beta$ the number of idle atoms left in the entanglement zone. $\tau_{\mathrm{trans}}$ and $\tau_{\mathrm{idle}}$, obtained by inverting Eq.~\eqref{eq:coherence}, express one atom transfer and one idle-atom exposure as an equivalent time of decoherence across all $n$ qubits.

The placer evaluates this cost incrementally. It keeps the waves $B$ of the moves already planned in the stage, and each new move either joins the compatible wave whose duration it increases least, or forms a new wave if none is compatible. A placement decision that adds the moves $V$ to the stage costs the marginal increase it causes,
\begin{equation}
\Delta T_{\mathrm{eqv}}(V) = T_{\mathrm{eqv}}(B \cup V) - T_{\mathrm{eqv}}(B),
\label{eq:delta}
\end{equation}
where $B \cup V$ denotes the waves after the moves in $V$ are inserted.

\paragraph{Idle-qubit evacuation}
\label{sec:idle}

A qubit in the entanglement zone that does not gate in stage $\ell$ suffers crosstalk from the global Rydberg pulse. Let $k_q$ be the number of two-qubit stages after $\ell$ and before $q$'s next two-qubit gate (or the end of the circuit if there is none). The placer visits such qubits in increasing order of $k_q$, with qubits that have no remaining gate last, so that those returning soonest get the most accessible storage traps. For each qubit, let $v$ be the move to the free storage trap with the smallest $\Delta T_{\mathrm{eqv}}(v)$ (Eq.~\eqref{eq:delta}). A qubit is evacuated along $v$ only if staying costs more than leaving,
\begin{equation}
T_{\mathrm{stay}} = k_q\,\tau_{\mathrm{idle}} > T_{\mathrm{leave}} = \Delta T_{\mathrm{eqv}}(v).
\label{eq:evac}
\end{equation}

\paragraph{Rydberg pair selection}
\label{sec:Rydberg-pair}
Next, the placer assigns each two-qubit gate a Rydberg pair site. If the two atoms of a gate already occupy the two traps of one pair, neither atom moves. If only one atom already sits in a Rydberg pair whose other trap is free, that atom stays and only its partner moves into the free trap.

The placer then handles the remaining gates in decreasing order of the distance between their two atoms. For each gate, it shortlists the $K = 4$ free pairs with the smallest combined travel distance. It then considers each candidate in both orientations and chooses the one with the smallest $\Delta T_{\mathrm{eqv}}(v_1, v_2)$ (Eq.~\eqref{eq:delta}), where $v_1$ and $v_2$ are the moves that bring the two atoms into the pair's two traps.

\subsection{Qubit Routing}
\label{sec:routing}

\subsubsection{Partial-progress routing}
\label{sec:ppr}

The router must realize all the moves $V$ implied by the next stage's placement without violating any of the four AOD rules (Sec.~\ref{sec:aod-model}). Algorithm~\ref{alg:ppr} gives the full procedure for a single stage, and Fig.~\ref{fig:ppr} compares it with prior work's routers.

\begin{algorithm}[t]
\caption{Partial-Progress Routing}
\label{alg:ppr}
\begin{algorithmic}[1]
\Statex \mbox{$V$: moves whose atoms have not been picked up}
\Statex \mbox{$S$: moves selected to finish in the next movement round}
\Statex \mbox{$W$: partial moves that stop at waypoints}
\Statex \mbox{$H$: remaining moves of held atoms (zero-length if arrived)}
\Function{RouteStage}{$V$}
  \State $H \gets \emptyset$
  \While{$V \cup H \ne \emptyset$}
    \State $S \gets \textsc{SelectWave}(V, H)$
    \State $W \gets \emptyset$
    \ForAll{$v \in H \setminus S$, then $v \in V \setminus S$}
      \State $w \gets \textsc{Waypoint}(v, S \cup W)$
      \If{$w \ne \bot$} $W \gets W \cup \{v_{\to w}\}$
      \ElsIf{$v \in H$} $W \gets W \cup \{v_{\to v}\}$
      \EndIf
    \EndFor
    \State $\textsc{PickUp}(S, W)$, $\textsc{Move}(S, W)$
    \State $V \gets V \setminus (S \cup W)$
    \State $H \gets (H \setminus (S \cup W))~  \cup$ remainders of $(S \cup W)$
    \State $H \gets \textsc{Release}(H)$
  \EndWhile
\EndFunction
\end{algorithmic}
\end{algorithm}

\begin{figure*}[!t]
\centering
\includegraphics[width=\textwidth]{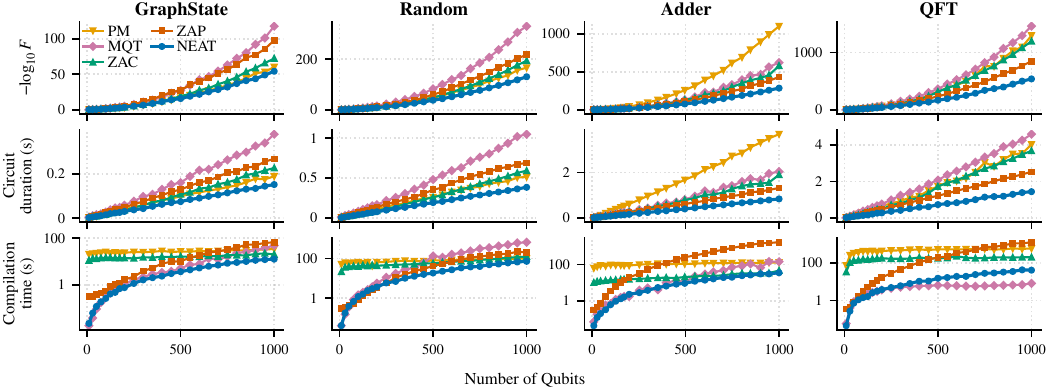}
\caption{Scalability study. Top: negative log-fidelity $-\log_{10} F$ (lower is better). Middle: circuit duration. Bottom: compilation time (log scale).}
\label{fig:scal}
\end{figure*}

\textsc{SelectWave} first selects the wave $S$ of moves that can finish in the next movement round. A move can join $S$ only if it is AOD-compatible with every held atom and every move already in $S$, and if its target trap is empty. A move whose target is occupied is also considered if the occupant's move is in $S$, because the occupant will have left by the time the move arrives. Such dependencies link the moves in $V$ into linear chains. For example, $v_1$ targets the trap that $v_2$ vacates, $v_2$ the trap that $v_3$ vacates, and so on, up to the \emph{free end}, a move whose target is empty. Since the placer moves each atom at most once per stage, into a free trap one at a time, the dependencies never close into a cycle. \textsc{SelectWave} visits the chains from their free ends, longest first. A move is considered only after the move it depends on has been admitted, and a chain closes at its first AOD-incompatible move, since neither it nor any move behind it can join $S$. The pass ends once every chain is closed or fully admitted.

The router then visits the blocked moves $v \in V \setminus S$ to make partial progress toward their targets. By Rules~1--3 (Sec.~\ref{sec:aod-model}), the positions to which $v$ can still move while staying compatible with every selected move form a rectangular region. $\textsc{Waypoint}(v, S \cup W)$ returns the point $w$ of this region closest to the target trap of $v$, or $\bot$ if it has no legal waypoint (Sec.~\ref{sec:aod-legality}). $W$ stores the leg $v_{\to w}$ from each such atom's current position to its waypoint. Once every blocked move has been visited, $\textsc{PickUp}(S, W)$ lifts the atoms of $S \cup W$ (Sec.~\ref{sec:aod-legality}). The moves in $S \cup W$ then run in parallel, and each atom's remaining move forms $H$. Finally, $\textsc{Release}(H)$ releases the atoms that have already arrived (Sec.~\ref{sec:aod-legality}).

Releasing atoms relaxes the AOD constraints, so more atoms can join the next wave. As many remainders in $H$ and moves in $V$ as can finish in the next movement round are then admitted. The waypoint search is then repeated, first for $H$, then for the blocked moves in $V$. \textsc{SelectWave} and \textsc{Waypoint} repeat until $V$ and $H$ are empty, at which point every atom has reached its final target trap.

\subsubsection{Legal AOD transport}
\label{sec:aod-legality}

\textsc{SelectWave} and \textsc{Waypoint} enforce Rules~1--3 (Sec.~\ref{sec:aod-model}) by construction, since they only admit moves that are compatible with every selected move. Rule~4 (ghost traps) and other routing technicalities are handled as follows.

\textsc{PickUp} lifts atoms in rounds. Each round lifts a maximal set of waiting atoms whose AOD lines create no ghost trap on an occupied site and stay at least $d_{\min}$ from every active line. The lifted atoms then slide off the trap grid so that more waiting atoms can be lifted cleanly in the next round. \textsc{Release} switches off one line of each arrived atom whose row or column carries no atom at a waypoint.

Routing can stall if every line of every arrived atom also holds an atom at a waypoint, at which point no atoms can be released. \textsc{Waypoint} prevents this jam by rejecting a waypoint (returning $\bot$) unless the AOD could afterwards be emptied by only releasing arrived atoms with a line that carries no atom at a waypoint. It also rejects a waypoint if it would cause both the row and column of an idle atom to contain atoms at waypoints, since this would leave a ghost trap on the idle atom that cannot be switched off, potentially dragging the idle atom along with the wave in a later movement round.

After compilation, the AOD verifier replays the instruction list and checks Rules~1--4 (Sec.~\ref{sec:aod-model}) on all active AOD lines, as well as other routing technicalities, \textit{e.g.,} that no two atoms occupied one position, only held atoms moved, every held atom was released on a trap, and every released atom had a line that can be switched off safely.

\section{Evaluation}
\label{sec:eval}

\subsection{Fidelity and Compilation Speed}
\label{sec:eval-compare}

We compare \ours{} against four SOTA compilers: PowerMove~\cite{powermove} (PM), MQT~\cite{stade2025search}, ZAC~\cite{zac}, and ZAP~\cite{zap}. Each compiler receives the same Qiskit-transpiled circuit, and its output is converted into a common instruction code and evaluated with the same fidelity model and hardware parameters (Sec.~\ref{sec:fidelity-model}). All compilers use their own array, except for MQT, which takes the array as input and uses ZAC's.

All experiments were run on an Apple M4 processor with 24~GB of memory. Each compilation time is normalized by a fixed calibration benchmark run before and after it. Every circuit compiled by \ours{} passes the built-in AOD verifier (Sec.~\ref{sec:aod-legality}) with zero violations.

\subsubsection{Standard benchmarks}
\label{sec:eval-qasmbench}

\begin{table}[!tbp]
\centering
\caption{Fidelity on the Standard Benchmarks}
\label{tab:qasmbench}
\footnotesize
% AUTO-GENERATED by paper/make_eval.py -- do not edit by hand.
\begin{tabular*}{\columnwidth}{@{\extracolsep{\fill}}lrrccccc@{}}
\toprule
 & & & \multicolumn{5}{c}{Fidelity $F$} \\
\cmidrule(l){4-8}
Circuit & $n$ & CZ & PM & MQT & ZAC & ZAP & NEAT \\
\midrule
qaoa & 6 & 36 & 0.716 & 0.727 & 0.733 & 0.745 & \textbf{0.746} \\
sat & 11 & 252 & 0.054 & 0.061 & 0.071 & 0.067 & \textbf{0.072} \\
bv & 14 & 13 & 0.837 & 0.843 & 0.850 & 0.852 & \textbf{0.860} \\
qram & 20 & 130 & 0.149 & 0.194 & 0.221 & 0.217 & \textbf{0.230} \\
swap\_test & 25 & 84 & 0.249 & 0.295 & 0.324 & 0.323 & \textbf{0.369} \\
ising & 26 & 50 & 0.625 & 0.619 & 0.619 & 0.651 & \textbf{0.671} \\
adder & 28 & 195 & 0.038 & 0.074 & 0.083 & 0.080 & \textbf{0.099} \\
dnn & 33 & 172 & 0.045 & 0.090 & 0.108 & 0.114 & \textbf{0.135} \\
qugan & 39 & 205 & 0.017 & 0.054 & 0.066 & 0.067 & \textbf{0.081} \\
ghz & 40 & 39 & 0.419 & 0.447 & 0.501 & 0.501 & \textbf{0.525} \\
\midrule
Geo.\ mean &  &  & 0.157 & 0.215 & 0.238 & 0.239 & \textbf{0.261} \\
\multicolumn{3}{@{}l}{Compilation time (s)} & 643.4 & 1.2 & 125.2 & 1.1 & \textbf{0.7} \\
\bottomrule
\end{tabular*}

\end{table}

We first compile 10 standard circuits of diverse structures and sizes from QASMBench~\cite{qasmbench} (Table~\ref{tab:qasmbench}). \ours{} achieves the highest fidelity on all 10 circuits, while compiling the suite fastest, showing consistent improvements across varied structures. PowerMove and ZAC are two to three orders of magnitude slower due to the simulated annealing they use for initial placement.

\subsubsection{Scalability}
\label{sec:eval-scal}

We compile 4 circuit families (GraphState, Random, Adder, and QFT) from 10 to 1{,}000 qubits and report the negative log-fidelity $-\log_{10} F$, circuit duration, and compilation time (Fig.~\ref{fig:scal}).

We observe that $-\log_{10} F$ correlates strongly with circuit duration, with the five compilers ranking in the same order by both metrics on every family's largest circuit. Each component of \ours{} helps shorten the circuit: partial-progress routing moves more atoms in parallel, transport-informed placement minimizes the incremental equivalent time of each decision, and \ours{} schedules gates to enhance parallelism and runs on an array that avoids AOD violations without extra moves. As a result, \ours{} has the shortest circuit duration on all 100 circuits and achieves the highest fidelity on 90 of them. In geometric mean over all circuits, \ours{} shortens the circuit duration by 2.0$\times$ (PowerMove), 2.3$\times$ (MQT), 1.6$\times$ (ZAC), and 1.6$\times$ (ZAP), or 1.9$\times$ across all four. Consequently, it lowers $-\log_{10} F$ by 1.6$\times$, 1.8$\times$, 1.4$\times$, and 1.4$\times$, respectively, or 1.5$\times$ across all four.

\begin{table*}[!tbp]
\centering
\caption{Placement and Objective Comparison Under a Common Router}
\label{tab:placer}
\scriptsize \setlength{\tabcolsep}{2.5pt}
% AUTO-GENERATED by paper/make_routing_eval.py -- do not edit by hand.
\begin{tabular*}{\textwidth}{@{\extracolsep{\fill}}l@{}c@{}|ccccc@{}c@{}|ccccc@{}c@{}|cccc@{}c@{}|cccc@{}}
\toprule
 & & \multicolumn{11}{c}{Placement} & & \multicolumn{9}{c}{Objective} \\
\cmidrule(lr){3-13}\cmidrule(l){15-23}
 & & \multicolumn{5}{c}{Avg.\ rank} & & \multicolumn{5}{c}{Regret (\%)} & & \multicolumn{4}{c}{Rank correlation} & & \multicolumn{4}{c}{Regret (\%)} \\
\cmidrule(lr){3-7}\cmidrule(lr){9-13}\cmidrule(lr){15-18}\cmidrule(l){20-23}
Family & & PM & MQT & ZAC & ZAP & NEAT &  & PM & MQT & ZAC & ZAP & NEAT &  & MQT & ZAC & ZAP & NEAT &  & MQT & ZAC & ZAP & NEAT \\
\midrule
QASMBench & & 1.33 & 0.51 & 0.63 & 0.62 & \textbf{0.45} &  & 18.8 & 2.0 & 2.4 & \textbf{1.1} & 1.5 &  & 0.79 & 0.70 & 0.76 & \textbf{0.81} &  & 0.3 & 0.2 & 0.2 & \textbf{0.1} \\
GraphState & & 2.03 & 1.70 & 2.08 & \textbf{1.16} & 1.40 &  & 30.4 & 25.8 & 29.2 & \textbf{6.1} & 6.5 &  & 0.53 & 0.41 & \textbf{0.61} & 0.58 &  & 4.1 & 4.3 & \textbf{2.4} & 2.8 \\
Random & & 2.79 & 3.10 & 2.14 & 1.63 & \textbf{0.32} &  & 15.5 & 18.4 & 14.9 & 10.8 & \textbf{0.5} &  & 0.48 & 0.17 & 0.31 & \textbf{0.50} &  & \textbf{0.7} & 14.7 & 10.4 & \textbf{0.7} \\
Adder & & 1.91 & 0.98 & 1.59 & 0.50 & \textbf{0.32} &  & 69.3 & 52.7 & 62.1 & 4.2 & \textbf{1.3} &  & 0.76 & 0.66 & 0.73 & \textbf{0.77} &  & 1.0 & 1.4 & 1.6 & \textbf{0.9} \\
QFT & & 2.04 & 1.38 & 1.39 & 0.67 & \textbf{0.14} &  & 172.7 & 86.6 & 80.7 & 8.0 & \textbf{0.6} &  & \textbf{0.61} & 0.45 & 0.44 & 0.59 &  & \textbf{1.6} & 7.4 & 8.0 & 2.4 \\
\midrule
Mean & & 2.11 & 1.67 & 1.69 & 0.95 & \textbf{0.54} &  & 67.1 & 41.9 & 42.7 & 6.7 & \textbf{2.1} &  & 0.61 & 0.45 & 0.54 & \textbf{0.63} &  & 1.7 & 6.3 & 5.1 & \textbf{1.6} \\
\bottomrule
\end{tabular*}

\end{table*}

\begin{table*}[!tbp]
\centering
\caption{Routing Comparison Under the Same Placements}
\label{tab:routing}
\scriptsize \setlength{\tabcolsep}{2.5pt} \renewcommand{\arraystretch}{1.1}
% AUTO-GENERATED by paper/make_routing_eval.py -- do not edit by hand.
\begin{tabular*}{\textwidth}{@{\extracolsep{\fill}}l@{}c@{}|ccc@{}c@{}|ccc@{}c@{}|ccc@{}c@{}|ccc@{}}
\toprule
Family &  & \multicolumn{3}{c}{Moves (incl.\ offsets)} &  & \multicolumn{3}{c}{Atoms per move (excl.\ offsets)} &  & \multicolumn{3}{c}{Moves per atom (excl.\ offsets)} &  & \multicolumn{3}{c}{Transport time (ms)} \\
\cmidrule(lr){3-5} \cmidrule(lr){7-9} \cmidrule(lr){11-13} \cmidrule(lr){15-17}
 &  & ZAC & \makecell{NEAT (w/o\\waypoints)} & \makecell{NEAT (with\\waypoints)} &  & ZAC & \makecell{NEAT (w/o\\waypoints)} & \makecell{NEAT (with\\waypoints)} &  & ZAC & \makecell{NEAT (w/o\\waypoints)} & \makecell{NEAT (with\\waypoints)} &  & ZAC & \makecell{NEAT (w/o\\waypoints)} & \makecell{NEAT (with\\waypoints)} \\
\midrule
QASMBench & & 98 & \textbf{92 ($-$6\%)} & \textbf{92 ($-$6\%)} &  & 1.31 & 1.27 ($-$3\%) & \textbf{1.61 ($+$23\%)} &  & 1.00 & 1.00 (0\%) & \textbf{1.26 ($+$26\%)} &  & 9.5 & 9.4 ($-$1\%) & \textbf{8.7 ($-$8\%)} \\
GraphState & & 676 & 366 ($-$46\%) & \textbf{347 ($-$49\%)} &  & 3.25 & 2.71 ($-$17\%) & \textbf{5.07 ($+$56\%)} &  & 1.00 & 1.00 (0\%) & \textbf{1.49 ($+$49\%)} &  & 61.5 & 58.7 ($-$5\%) & \textbf{43.8 ($-$29\%)} \\
Random & & 1{,}913 & 1{,}057 ($-$45\%) & \textbf{985 ($-$48\%)} &  & 4.32 & 3.60 ($-$17\%) & \textbf{6.76 ($+$56\%)} &  & 1.00 & 1.00 (0\%) & \textbf{1.53 ($+$53\%)} &  & 151.7 & 139.0 ($-$8\%) & \textbf{107.2 ($-$29\%)} \\
Adder & & 2{,}331 & \textbf{1{,}721 ($-$26\%)} & 1{,}739 ($-$25\%) &  & 1.79 & 1.72 ($-$4\%) & \textbf{2.39 ($+$34\%)} &  & 1.00 & 1.00 (0\%) & \textbf{1.34 ($+$34\%)} &  & 228.2 & 217.7 ($-$5\%) & \textbf{193.2 ($-$15\%)} \\
QFT & & 6{,}429 & \textbf{3{,}376 ($-$47\%)} & 3{,}537 ($-$45\%) &  & 2.50 & 2.50 (0\%) & \textbf{3.40 ($+$36\%)} &  & 1.00 & 1.00 (0\%) & \textbf{1.36 ($+$36\%)} &  & 447.3 & 334.5 ($-$25\%) & \textbf{318.2 ($-$29\%)} \\
\midrule
Geo.\ mean & & 1{,}588 & 968 ($-$39\%) & \textbf{954 ($-$40\%)} &  & 2.63 & 2.39 ($-$9\%) & \textbf{3.76 ($+$43\%)} &  & 1.00 & 1.00 (0\%) & \textbf{1.41 ($+$41\%)} &  & 134.7 & 121.0 ($-$10\%) & \textbf{101.9 ($-$24\%)} \\
\bottomrule
\end{tabular*}

\end{table*}

\ours{} also has the highest compilation efficiency. It is the fastest compiler on 66/100 circuits and compiles all 100 circuits in 21~minutes, achieving a geometric-mean speedup of 23.7$\times$ (PowerMove), 1.4$\times$ (MQT), 10.1$\times$ (ZAC), and 5.4$\times$ (ZAP), or 6.6$\times$ across all four. Its compilation time is also predictable. No circuit takes \ours{} longer than 72~s (random\_n1000), whereas every other compiler exceeds 100~s on between 14 (MQT) and 39 (PowerMove) circuits. This shows that our deterministic flow scales well with qubit count.

\subsection{Placement Analysis}
\label{sec:eval-components}

To isolate placement, we compare the five placers stage by stage. We take the 10 QASMBench circuits and the scalability circuits of 10 to 1{,}000 qubits, 110 in total, and sample up to 10 evenly spaced two-qubit stages from each. At every sampled stage, all five placers start from the same mapping and propose the placement for the next stage. \ours{}'s router routes every proposal, and we assess it by the fidelity loss of the routed stage (Eq.~\eqref{eq:fidelity}). An idle atom left in the entanglement zone is charged crosstalk for every two-qubit stage until its next gate, because leaving it there is a decision made at this stage. For each placer, we report its rank (0 for the best) and its regret (its extra loss over the best, in percent), averaged over the stages of each circuit, then over circuits, so that each circuit counts equally.

We also test the objectives of four compilers by letting them rank the same five candidates at every stage. \ours{} uses $T_{\mathrm{eqv}}$ (Eq.~\eqref{eq:stage-cost}). MQT sums the square root of the longest move in each compatible set. ZAC sums the square roots of the move distances. ZAP sums the move distances and adds a penalty of 1{,}000 (default) per pair of conflicting moves. PowerMove is excluded because its placer follows fixed rules. Since the objectives of MQT, ZAC, and ZAP do not model crosstalk, we remove it from both the fidelity loss and $T_{\mathrm{eqv}}$ here, judging every objective only on how well it evaluates the moves. We report each objective's Kendall rank correlation with the fidelity loss and the regret of the candidate it ranks first, averaged as above.

Table~\ref{tab:placer} compares the placers (left) and the objectives (right). Among the placers, \ours{} has the lowest average rank (0.54) and regret (2.1\%), showing that it is able to find high-quality placements consistently. Among the objectives, $T_{\mathrm{eqv}}$ has the highest rank correlation (0.63) and the lowest regret (1.6\%), making it the most reliable objective even with its crosstalk term removed. The other compilers, however, come much closer to \ours{} on the objectives than on the placers. Their placers have regrets of 6.7\% to 67.1\% (3.1 to 31.3 times \ours{}'s), whereas their objectives have regrets of only 1.7\% to 6.3\% (1.1 to 4.0 times $T_{\mathrm{eqv}}$'s). This suggests that the different heuristics vary less in how accurately they rank given candidates than in how well they guide the placer toward high-quality placements.

\subsection{Routing Analysis}
\label{sec:eval-routing}

Prior routers share a common mechanism: they group moves by AOD compatibility, then pick up, move, and drop off each group as a unit (Sec.~\ref{sec:related}). We therefore compare ZAC's router, as an example, with \ours{}'s (with and without waypoints) (Table~\ref{tab:routing}), routing the same 110 circuits as Table~\ref{tab:placer} in full using \ours{} placements on its array. Each metric is computed over a whole circuit, then averaged geometrically over circuits so that large circuits do not dominate.

With or without waypoints, \ours{}'s router needs about 40\% fewer moves (including offsets) than ZAC's. ZAC's router picks up atoms row by row, with an offset move between rows to avoid ghost traps, whereas \ours{}'s \textsc{PickUp} (Sec.~\ref{sec:aod-legality}) exploits the staggered array (Sec.~\ref{sec:array}) to lift every cleanly liftable atom in each pick-up round. Together with the rest of \ours{}'s router, this lowers the transport time by 10\% compared with ZAC's, even without waypoints.

With waypoints, each move (excluding offsets) carries 43\% more atoms than ZAC's and 57\% more than it does without waypoints, because a conflicting atom can still join a wave and stop at an intermediate waypoint. Each atom stops at an average of 0.41 waypoints, whereas the other two routers move every atom exactly once. Together, the full version of \ours{}'s router lowers the total transport time by 24\% compared with that of ZAC's router.

\section{Conclusion}
\label{sec:conclusion}

We presented \ours{}, a zoned neutral-atom compiler that evaluates each placement decision by its incremental equivalent time and routes qubits in parallel through partial progress. Together with its staggered array and parallelism-enhancing schedule, \ours{} achieves substantially higher fidelity, shorter circuit duration, and better compilation efficiency than state-of-the-art compilers. Isolated analyses further show that both its placer and router outperform those of prior compilers. We therefore believe \ours{} will improve the performance of current neutral-atom quantum computers and help pave the way toward the early fault-tolerant regime of quantum computing at scale.

\section*{Acknowledgment}

The authors thank Chen Huang for her valuable comments.

\balance
\bibliographystyle{IEEEtran}
\bibliography{references}

\end{document}